\documentstyle[epsfig,preprint,tighten,prc,aps]{revtex}

\def\lsim{\mathrel{\rlap{\lower4pt\hbox{\hskip1pt$\sim$}}
    \raise1pt\hbox{$<$}}}         

\begin{document}
\draft

\title{The reactions $pn \rightarrow d\omega$ and 
$pn \rightarrow d\phi$ near threshold}

\author{K. Nakayama$^{a,b}$, J. Haidenbauer$^b$, and J. Speth$^b$}

\address{
$^a$Department of Physics and Astronomy, University of Georgia, Athens, GA 
30602, USA \\
$^b$Institut f\"{u}r Kernphysik, Forschungszentrum J\"{u}lich
GmbH, D--52425 J\"{u}lich, Germany 
}

\maketitle

\begin{abstract}
The reactions $pn \rightarrow d\omega$ and $pn \rightarrow d\phi$ 
are studied within a relativistic meson-exchange model
of hadronic interactions. Predictions for the total cross sections
and for the angular distributions of the vector mesons are
presented. The resulting cross sections near threshold
are around 10 - 30 $\mu$b for $pn \rightarrow d\omega$ and 200 - 250 nb
for $pn \rightarrow d\phi$. A moderate deviation of the cross section 
ratio $\sigma_{pn\rightarrow d\phi} / \sigma_{pn\rightarrow d\omega}$
from that of the Okubo-Zweig-Iizuka rule is predicted.


\end{abstract}
\newpage

 \section{Introduction}

The possibility that there might be a rather significant strangeness
content in the nucleon, as indicated by the analysis of the
$\pi$-nucleon $\Sigma$ term \cite{Don86} and the lepton 
deep-inelastic scattering data by the EMC and successor 
experiments \cite{EMC88}, has triggered a wealth of
experimental and theoretical investigations on this topic over the
last decade. In this context, hadronic reactions at low and medium 
energy play an important role. Here one expects that the presence
of strangeness in the nucleons would manifest itself in counting
rates that significantly exceed predictions based on the Okubo-Zweig-Iizuka
(OZI) rule \cite{OZI,Lip76}. Indeed, the experiments on
antiproton-proton ($\bar pp$) annihilation at rest at the LEAR facility 
at CERN revealed a strong violation of the OZI rule for various 
decay channels involving the $\phi$ meson \cite{Ast,CrB,Obe}.
Substantial deviations from the OZI rule were also found
in the reactions $pd \rightarrow ^3$He$\phi$ \cite{Wur96} 
and $\bar p p\rightarrow \phi\phi$ \cite{Ber95}. 
More recently the DISTO collaboration reported on an experiment on
exclusive production of $\phi$ and $\omega$ mesons in proton-proton
($pp$) collisions \cite{Disto}. In this measurement a cross section 
ratio of
about a factor 10 larger than the estimate from the OZI
rule was observed (at a cms energy of about 80 MeV above the
$\phi$ production threshold). Further experimental investigations of 
the reactions $pp\rightarrow pp\phi$ and $pp\rightarrow pp\omega$,
as well as $pn\rightarrow d\phi$ and $pn\rightarrow d\omega$,
closer to their thresholds, are planned for the near future at
the COSY facility in J\"ulich \cite{Cosy1,Cosy2}. 

Recently we have developed a model for the reactions
$pp \rightarrow pp\omega$ and $pp \rightarrow pp\phi$ 
within the conventional meson-exchange picture \cite{Nak1,Nak2}. 
In this model meson production is described by the two mechanisms
depicted in Fig.~\ref{fig1}, namely the nucleonic current and
the $\pi\rho v$ ($v = \omega, \phi$) meson exchange current. 
Those two processes were identified as providing the dominant
contributions to these vector-meson production reactions. 
We employed this model for a
combined analysis of close-to-threshold $\phi$ and $\omega$
production data with special emphasis on the apparent OZI rule violation 
reported by the DISTO collaboration. It turned out that the data
do not require a large $NN\phi$ coupling constant. Indeed, the range
of values for $g_{NN\phi}$ extracted from this analysis was found
to be compatible with the OZI value \cite{Nak2} - which might be 
interpreted as an indication that there is no need for introducing 
an $\bar ss$ component into the nucleon in order to describe those data. 
On the other hand, it was necessary to introduce a violation of the OZI
rule at the $\pi\rho\phi$ vertex in the meson-exchange current - which, 
in turn, provided the enhancement of the $\phi$ production cross 
section over the OZI estimation as seen in the DISTO data. 

In the present paper we apply this model to the reactions
$pn \rightarrow d\omega$ and $pn \rightarrow d\phi$. 
First of all, it is desirable to have quantitative results for those
reaction channels. 
In this context we would like to emphasize that all our model 
parameters have already been fixed in the previous study \cite{Nak2}, 
where they were constrained by requiring a consistent description of the 
reactions $pp \rightarrow pp\omega$ and $pp \rightarrow pp\phi$. 
Therefore we are able to provide genuine predictions for observables
for $pn \rightarrow d\omega$ and $pn \rightarrow d\phi$. Such
predictions should be very useful for corresponding measurements
that are planned at the COSY facility in J\"ulich.
Furthermore they can be used for testing the reliability of our
model when data become available in the future. 

Secondly, in our previous studies \cite{Nak1,Nak2} 
we found that the angular distribution of the produced vector mesons
is very sensitive to the production mechanisms. 
We showed that this angular distribution enables us to
determine both the relative importance of the basic reaction mechanisms 
(cf. Fig.~\ref{fig1}) and their absolute magnitudes. In fact, it was
this special feature that enabled us to extract the $\phi NN$ coupling 
constant from the analysis in Ref. \cite{Nak2}. 
Thus it is interesting to see whether a similar signature is also
present in the reactions 
$pn \rightarrow d\omega$ and $pn \rightarrow d\phi$ and, specifically,
whether these reactions are even better suited for discriminating
between the different production mechanisms. 

Finally, 
in the aftermath of the LEAR experiments, Ellis et al. \cite{Ell95} 
proposed a qualitative model in which it is assumed that the nucleon 
contains, besides $u$ and $d$ quarks, an intrinsic $\bar s s$ 
component. In this model $\phi$ production proceeds via a ``shake-out'' 
or a ``rearrangement'' of the $s$ and $\bar s$ quarks in the
nucleon. Assuming, furthermore, that the $\bar ss$ component in
the nucleon is polarized introduces selection rules for the
various $\phi$ production channels and allows a qualitative 
interpretation of the state dependence shown by the $\bar p p$
annihilation data.
Recently this picture has also been used for some qualitative
estimates for the $\phi$ production in $NN$ collisions \cite{Ell99}. 
It is interesting to explore whether the predictions of our 
conventional model differ from the results that would follow from 
the picture of Ellis et al. 

The paper is structured in the following way: In Section II we give
a review of our relativistic meson-exchange model for
vector-meson production. In Section III we present and discuss
our results for the reactions $pn \rightarrow d\omega$ and 
$pn \rightarrow d\phi$. The paper concludes with a short summary.

 \section{The model}

We study the reaction $pn \rightarrow d v$ ($v=\omega, \phi$) 
in a DWBA approach \cite{Nak1,Nak2}. The transition amplitude is given by 
\begin{equation}
M^\mu = < \psi_d|J^\mu|\phi_i> \ , 
\label{DWBA}
\end{equation}
where $\psi_d$ is the deuteron wave function and $\phi_i$ is the
four-component unperturbed $NN$ wave function in the initial state. 
We take the deuteron wave function of the model Bonn B as 
defined in Table A.1 of Ref.~\cite{MHE87}. (Note that this model has been 
also used to generate the $pp$ final state interaction in our previous
study \cite{Nak1,Nak2} of the $pp\rightarrow ppv$ reaction.) 
$J^\mu$ is the current describing the vector-meson production process. 
In our model $J^\mu$ is given by the sum of the nucleonic 
and $v\rho\pi$ meson-exchange currents, 
$J^\mu = J^\mu_{nuc}  + J^\mu_{mec}$, as illustrated diagrammatically in 
Fig.~\ref{fig1}. In principle there are, of course, many more 
meson-exchange currents that could contribute. Their relevance 
was thoroughly investigated in \cite{Nak2} and it was found that all
other exchange currents yield contributions that are much smaller
than those of the $v\rho\pi$ meson-exchange current.

Explicitly, the nucleonic current is defined as
\begin{equation}
J^\mu_{nuc} = \sum_{j=1,2}\left ( \Gamma^\mu_j iS_j U + U iS_j \Gamma^\mu_j 
\right ) \ ,
\label{nuc_cur}
\end{equation}
with $\Gamma^\mu_j$ denoting the $NNv$ vertex and $S_j$ the nucleon (Feynman) 
propagator for nucleon $j$. The summation runs over the two interacting 
nucleons, 1 and 2. $U$ stands for the meson-exchange $NN$ potential. It is, in 
principle, identical to the driving potential \cite{MHE87} used in the 
calculation of the deuteron wave function $\psi_d$ in Eq.~(\ref{DWBA}), 
except that here meson retardation effects are retained following the 
Feynman prescription. Eq.~(\ref{nuc_cur}) is illustrated in Fig.~\ref{fig1}a.

The structure of the $NNv$ vertex, $\Gamma^\mu$ (the subscript $j=1,2$ is 
omitted), required in Eq.~(\ref{nuc_cur}) for the production is obtained from
the Lagrangian density
\begin{equation}
{\cal L}(x) =  - \bar\Psi(x) \left( g_{NNv}
       [\gamma_\mu -{\kappa_v\over 2m_N}\sigma_{\mu\nu}\partial^\nu ] V^\mu(x)
             \right) \Psi(x) \ ,
\label{NNv}
\end{equation}
where $\Psi(x)$ and $V^\mu(x)$ stand for the nucleon and vector-meson fields,
respectively. $g_{NNv}$ denotes the vector coupling constant and $\kappa_v 
\equiv f_{NNv}/g_{NNv}$, with $f_{NNv}$ the tensor coupling constant. 
$m_N$ denotes the nucleon mass.

As in most meson-exchange models of interactions, each hadronic 
vertex is furnished with a form factor in order to account for, among other 
things, the composite nature of the hadrons involved. In this spirit
the $NNv$ vertex obtained from the above Lagrangian 
is multiplied by a form factor. 
The theoretical understanding of this form factor is beyond the 
scope of the present paper; we assume it to be of the form
\begin{equation}
F_N(l^2) = \frac{\Lambda_N^4} {\Lambda_N^4 + (l^2-m_N^2)^2}  \ ,
\label{formfactorN}
\end{equation}
where $l^2$ denotes the four-momentum squared of either the incoming or
outgoing off-shell nucleon. It is normalized to unity when the nucleon is
on its mass-shell, $l^2 = m_N^2$.

The $v\rho\pi$ vertex required for constructing the meson-exchange current,
$J^\mu_{mec}$ (Fig.~\ref{fig1}b), is derived from the Lagrangian density
\begin{equation}
{\cal L}_{v\rho\pi}(x) = \frac{g_{v\rho\pi}} {\sqrt{m_v m_\rho}}
\varepsilon_{\alpha\beta\nu\mu} \partial^\alpha \vec \rho^\beta(x) \cdot
\partial^\nu \vec \pi(x) V^\mu(x) \ ,
\label{pirhoomega}
\end{equation}
where $\varepsilon_{\alpha\beta\nu\mu}$ denotes the Levi-Civita antisymmetric
tensor with $\varepsilon_{0123}=-1$. The $v\rho\pi$ vertex obtained from the 
above Lagrangian is multiplied by a form factor which is taken to be of the
form
\begin{equation}
F_{v\rho\pi}(q_\pi^2, q_\rho^2) = 
\left ( \frac{\Lambda_M^2 - m_\pi^2} {\Lambda_M^2 - q_\pi^2} \right )
\left ( \frac{\Lambda_M^2 - x m_\rho^2} {\Lambda_M^2 - q_\rho^2} \right ) \ .
\label{formfactorM}
\end{equation}
It is normalized to unity at $q_\pi^2 = m_\pi^2$ and $q_\rho^2 = x m_\rho^2$.
The parameter $x$ (= 0 or 1) is introduced in order to allow for different
normalization points in order to take into account that the 
values of the $v\rho\pi$ coupling constants that we employ are extracted
at different kinematics \cite{Nak2}. $g_{\phi\rho\pi}$ is determined at 
$q^2_\rho = m^2_\rho$ and $q^2_\pi = m^2_\pi$, whereas 
$g_{\omega\rho\pi}$ is extracted at $q^2_\rho = 0$ and $q^2_\pi = m^2_\pi$. 
Accordingly, we use the normalization $x=1$ for the form factor at the
former vertex and $x=0$ for the latter. 

The meson-exchange current is then given by
\begin{equation}
J^\mu_{mec} = [\Gamma^\alpha_{NN\rho}(q_\rho)]_1 iD_{\alpha\beta}(q_\rho)
              \Gamma^{\beta\mu}_{v\rho\pi}(q_\rho, q_\pi, k_\omega)
              i\Delta(q_\pi) [\Gamma_{NN\pi}(q_\pi)]_2   +  
(1\leftrightarrow 2) \ ,
\label{mec_cur}
\end{equation}
where $D_{\alpha\beta}(q_\rho)$ and $\Delta(q_\pi)$ stand for the $\rho$- and 
$\pi$-meson (Feynman) propagators, respectively. The vertices $\Gamma$ 
involved are self-explanatory. Both the $NN\rho$ and $NN\pi$ vertices,
$\Gamma^\alpha_{NN\rho}$ and $\Gamma_{NN\pi}$, are taken consistently with
the $NN$ potential used to generate the deuteron wave function in 
Eq.~(\ref{DWBA}).

Our model described above contains five parameters to be fixed for a given 
vector-meson produced: two for the mesonic current (the coupling constant 
$g_{v\rho\pi}$ and the cutoff parameter $\Lambda_M$) and three for the 
nucleonic current (the coupling constants $g_{NNv}$ and $f_{NNv}$, and the 
cutoff parameter $\Lambda_N$). 

The coupling constant $g_{\phi\rho\pi}$ was extracted from the 
measured decay width ($\phi \to \rho + \pi$) and $g_{\phi\rho\pi}$ 
from the radiative
decay width ($\omega \to \gamma + \pi$) assuming vector meson
dominance. All other parameters were fixed in a 
combined analysis of all available
near-threshold data on $\phi$- and $\omega$-meson production in $pp$ 
collisons \cite{Nak2}. Besides using the data from both reactions, we 
also assumed relations between corresponding parameters in the production
amplitudes. Specifically, we assumed that the form factors at the 
$\phi\rho\pi$ and $\omega\rho\pi$ vertices are the same. Likewise, 
we assumed that the form factor at the meson production vertices in the
nucleonic current are the same. The $NN\omega$ coupling constant is fixed
to the SU(3) value, $g_{NN\omega}$ = 9, based on $g_{NN\rho}$ of 
Ref.~\cite{rhocoup}. Furthermore we assumed that $\kappa_\phi = 
\kappa_\omega \equiv \kappa$, as also suggested by SU(3) symmetry,
with $-0.5 \leq \kappa \leq 0.5$. 

Details of our strategy for fixing the model parameters are outlined 
in Ref.~\cite{Nak2}. Here we only want to mention that the lack of a more 
complete set of data prevented us from achieving a unique solution.
Rather, we arrived at four different model solutions that all provide
a comparably good description of the data but differ in the value of
the nucleon cutoff mass $\Lambda_N$, the value of $g_{NN\phi}$ and the 
value of $\kappa$, cf.
Table \ref{tab1}. We shall employ all these model solutions in the present 
investigation. 

Finally, we want to mention that the initial state interaction (ISI) 
is not taken into account explicitly in our model. Rather, its effect is
accounted for effectively via an appropriate adjustment of the
(phenomenological) form factors at the hadronic vertices, cf. 
the corresponding statements in Ref. \cite{Nak2}. Admittedly, this 
procedure leads to a certain degree of uncertainty of our predictions
because the ISI in the reactions $pn\rightarrow dv$ 
and $pp\rightarrow ppv$, respectively, is governed by different
partial waves (different isospin states). On the other hand, we don't 
have at present any better and more solid alternative for
dealing with the ISI interaction. There are no $NN$ interaction models
in the literature that are still valid in the energy range relevant 
for $\omega$ and $\phi$ production 
for an explicit treatment of ISI effects. 
Furthermore, the qualitative prescription proposed in 
Ref. \cite{HaKa}, based on phase shifts and inelasticity parameters,
can't be used either because there is no reliable phase shift analysis 
beyond 1.3 GeV for the isospin $T$ = 0 channel \cite{Arndt} 
that we need here. 

\section{Results and discussion}

The data that were used in Ref.~\cite{Nak2} to fix the free parameters
of our model consisted of some near-threshold total cross sections
for the reaction $pp\rightarrow pp\omega$ \cite{Hib99} and of the total
cross section, as well as the angular distribution of the $\phi$ at
$T_{lab}$ = 2.85 GeV for the reaction 
$pp\rightarrow pp\phi$ \cite{Disto}. Naturally, the more selective
experimental information available for the latter reaction (angular
distributions) put also
a stronger constraint on the model parameters and, as a
consequence, the predictions for $pn\rightarrow d\phi$ for the
four parameter sets in Table \ref{tab1} don't differ much. As a
matter of fact, the general features of the results are practically 
identical and therefore we show only the predictions for one of the sets
(set 1 of Table \ref{tab1}) in Fig.~\ref{stphi}. Values for the
total cross section at a few excess energies $Q$ are, however, compiled 
in Table~\ref{tab2} for all parameter sets. 
It is evident from Fig.~\ref{stphi} that the reaction $pn\rightarrow d\phi$
is strongly dominated by the meson exchange current. The relative
magnitude of the contribution from the nucleonic current is even 
smaller than in case of $pp\rightarrow pp\phi$; it was about 18 \% 
there, here it is a mere 3 \%.
The stronger suppression of the nucleonic current in the reaction 
$pn\rightarrow d\phi$ is due to the fact that
the meson exchange current is weighted by a larger isospin coefficient
({\rm -3} for $pn\rightarrow d\phi$ as compared to {\rm +1} in the 
case $pp\rightarrow pp\phi$). The nucleonic current, however, has
contributions from the exchange of iso-scalar as well as iso-vector
mesons. Consequently, only the latter are multiplied
by a larger isospin coefficient. Furthermore, since the isospin factor
changes sign, there are now also cancelations between the meson exchanges 
that contribute to the nucleonic current.
As in the case of $pp\rightarrow pp\phi$, there is a destructive interference
between the nucleonic current and the meson exchange current which, 
however, is now much less pronounced. Nonetheless, the predicted
angular distributions of the emitted $\phi$ meson, which are fairly 
flat, show some downward bending at very forward and backward angles
as a consequence of this destructive interference between the two production
mechanisms, cf. Fig.~\ref{sdphi}. 
For higher energies (Fig.~\ref{sdphi}, $Q=100 MeV$) the suppression of the 
angular distribution at forward and backward angles becomes more
pronounced and it should be possible to verify experimentally 
whether the angular distribution does indeed have such a structure.
In general there is not much difference between the
angular distributions resulting from the four parameter
sets, except that those with $\kappa$ = 0.5 yield a total cross
section which is about 20 \% larger, cf. also Table~\ref{tab2}. 

Predictions of total cross sections for the reaction $pn \rightarrow d\omega$ 
for the four parameter sets are shown in Fig.~\ref{stome}.
Also here we observe that the meson exchange current
is the dominant production mechanism. Its relative importance is
likewise increased compared to $pp\rightarrow pp\omega$. Indeed, now
the meson exchange current dominates for all four parameter sets,
whereas in the reaction $pp\rightarrow pp\omega$ it is the nucleonic 
current that provides the dominant contribution for the parameter 
sets 2 and 4. 
Still, in all cases, the nucleonic current is large enough to introduce
sizable interference effects. Consequently, we obtain a much stronger
variation of the predicted total cross sections for $pn\rightarrow d\omega$
than for $pn\rightarrow d\phi$. 
The results vary by a factor of roughly 3, as can be seen from 
Fig.~\ref{stome} and also from Table~\ref{tab2}. 
Note that there are particularly strong interference effects in case of
parameter set 2 (Fig.~\ref{stome}b), where the resulting total cross section
is significantly smaller than the individual contributions. 
As we shall discuss later, 
the stronger variations in the $\omega$ production cross section have
also consequences for the cross section ratio $R_{\phi/\omega} \equiv
\sigma_{pn\rightarrow d\phi} / \sigma_{pn\rightarrow d\omega}$
relevant for the comparison with the 
corresponding ratio that follows from the OZI rule.

We observe that for $Q \leq$ 20 MeV, effects from the finite width
of the $\omega$ meson become important. These effects are
included in our calculation by folding the calculated cross
section with the Breit-Wigner mass distribution of the $\omega$
meson and can be seen in Fig.~\ref{stome}. 

Predictions for the angular distribution of the $\omega$ meson
are displayed in Fig.~\ref{sdome}. Naturally, here we see also 
larger variations between the predictions for the different
parameter sets. But, since the meson exchange current is the dominant
production mechanism for all sets, all the angular distributions show 
qualitatively a similar structure, i.e. they are rather flat at the 
lower energy ($Q$ = 30 MeV) and exhibit a more-or-less strong downward 
bending at forward and backward angles at the higher energy 
($Q$ = 100 MeV). We would like to emphasize, however, that this
behavior differs from the one in the reaction 
$pp\rightarrow pp\omega$. In the latter reaction a clear enhancement 
of the angular distribution in forward and backward direction 
was seen in the DISTO experiment \cite{Disto} (at $T_{lab}$ = 2.85 MeV), 
and such an enhancement for the reaction $pp \rightarrow pp\omega$ is 
also predicted by our model (parameter sets 2 and 4) - 
even for energies closer to the $\omega$ production threshold. 

As a general feature, let us also mention that the 
cross sections for the reactions $pn\rightarrow dv$ are
significantly larger than the ones for $pp\rightarrow ppv$ near
threshold. Apart from the isospin factors involved, 
such a behavior can be understood in terms of phase space
arguments (the two body phase space opens up more rapidly close 
to threshold) and is well-known from other meson production 
reactions (see, e.g., Refs. \cite{Mey97,Cal98}). 

We now turn our attention to the cross section ratio 
$R_{\phi/\omega} =
\sigma_{pn\rightarrow d\phi} / \sigma_{pn\rightarrow d\omega}$
for a comparison with the prediction that follows from the naive 
application of the OZI rule \cite{Lip76}. The adjective ``naive'' 
refers to the fact that the cross section ratio is simply equated to
the square of the ratio of the relevant coupling constants at the 
$\phi$ and $\omega$ production vertices. The OZI rule plus SU(3) 
imply that these coupling constants are 
proportional to $\sin \alpha_v $ and $\cos \alpha_v $, respectively, 
where $\alpha_v$ stands for the deviation of the $\phi$-$\omega$ 
mixing angle from its ideal mixing value of $\theta_{(ideal)}=35.3^o$. 
Thus the OZI estimate for the
cross section ratio is simply given by $R_{\phi / \omega} = 
\tan ^2\alpha_v$ \cite{Lip76}. With $\alpha_v=3.7^o$ as extracted 
from the quadratic mass formula, this yields a value of 
$R_{\phi / \omega} = 4.2 \times 10^{-3}$. 

Our results for $R_{\phi / \omega}$ are presented in Fig.~\ref{stratio}. 
The cross sections are evaluated at the same excess energy in order to
minimize effects from differences in the phase space. We see that
the largest values are predicted for excess energies close to threshold.
E.g., at $Q \approx$ 10 MeV we get values between $32 \times 10^{-3}$ (set 2)
and $9 \times 10^{-3}$ (set 3). Thus, compared to the ratio that
follows from the naive OZI rule, $R_{\phi / \omega}
= 4.2 \times 10^{-3}$, our models yield a moderate deviation of at most a 
factor of 8. At higher energies, $Q$ = 80 $\sim$ 100 MeV, the enhancement
over the OZI estimate decreases to a factor of around 3 or so. 
 
So, what is the origin of this enhancement over the OZI estimate as 
predicted by our model? 
As already mentioned in the Introduction, our model involves
an explicit violation of the OZI rule at the $\pi\rho\phi$ 
vertex in the meson exchange current. As discussed in detail 
in Ref.~\cite{Nak2}, it had to be introduced in order to achieve a 
simultaneous and consistent description within our model of the 
available data on the reactions $pp\rightarrow pp\omega$ and
$pp\rightarrow pp\phi$, particularly of the energy dependence of the
total $\omega$ production cross section and the angular distribution 
of the $\phi$ meson. This explicit OZI violation, in terms of the 
$\pi\rho\phi$ and $\pi\rho\omega$ coupling constants used, 
suggests an enhancement of around 3 in the cross section ratio.  
With regard to the nucleonic current, the employed $NN\omega$ and 
$NN\phi$ coupling constants lead to results that exceed the OZI 
value only in one case, namely for parameter set 3, cf. Table~\ref{tab1}. 
The corresponding enhancement factor for the cross section ratio 
amounts to around 2. 

It is evident from Fig.~\ref{stratio} that the cross section ratios
resulting from the full model calculation differ significantly from
those values implied by the employed coupling constants. 
Obviously dynamical effects such as interferences, energy dependence 
of the production amplitudes, etc. play a rather important role here. 
For example, parameter set 3 - which is the only model where
the $NN\phi$ coupling exceeds the OZI value - yields the smallest
enhancement over the naive OZI estimate among all models considered,
cf. the dash-dotted curve in Fig.~\ref{stratio}. Furthermore, 
the particularly large enhancement over the OZI estimate found for parameter 
set 2 (especially at small excess energies) is, in fact, 
due to a suppression of the $\omega$ production cross section (resulting 
from strong interference effects as pointed out above) and not 
caused by an enhancement in the $\phi$ production, cf. Table~\ref{tab2}.
Apparently fairly large deviations from the naive OZI prediction can 
be generated by dynamical effects within the conventional picture: 
i.e., without introducing any ''exotic`` mechanisms. 
Consequently, one should be very cautious in drawing direct conclusions 
on the strangeness content in the nucleon from such cross section ratios.
The behavior of the 
cross section ratio as the excess energy approaches zero is due to the 
finite width of the $\omega$-meson which prevents the $\omega$-meson 
production cross section from decreasing rapidly as it does in the case of 
$\phi$-meson, cf. Figs.~\ref{stphi},\ref{stome}.

Evidently, the enhancement of a factor of 3 or so over the OZI estimate
at higher excess energies in Fig.~\ref{stratio} is much smaller than the 
enhancement over the OZI estimate of about a factor 10 found for  
$\sigma_{pp\rightarrow pp\phi} / \sigma_{pp\rightarrow pp\omega}$ 
by the DISTO collaboration \cite{Disto} at an excess energy 
$Q \approx$ 80 MeV of the $\phi$. However it is important to
realize that their measurement was done at a fixed incident beam
energy ($T_{lab}$ = 2.85 GeV); therefore the corresponding
excess energy of the produced $\omega$ is already 319 MeV. Although
corrections for the differences in the available phase space were 
obviously applied when extracting the above result, there are other
effects that influence the ratio such as the energy dependence of the
production amplitude, the onset of higher partial waves, etc.,
which one cannot correct for easily. Therefore it is possible that
the actual deviation from the naive OZI rule in the reactions 
$pp\rightarrow ppv$ is also smaller. Indeed our results for
$\sigma_{pp\rightarrow pp\phi} / \sigma_{pp\rightarrow pp\omega}$, 
shown in Fig.~\ref{ppratio}, support this conjecture. We see here that  
the results are very similar to the ones for 
$\sigma_{pn\rightarrow d\phi} / \sigma_{pn\rightarrow d\omega}$.
Specifically, at the excess energy of $Q$ = 80 MeV the enhancement
over the OZI estimate is also only around a factor 3. Thus 
it would be interesting to perform a measurement of the ratio 
$\sigma_{pp\rightarrow pp\phi} / \sigma_{pp\rightarrow pp\omega}$ 
at the same (or at least similar) excess energies. 
 
Finally, let us comment on a recent work by Ellis et al. \cite{Ell99}.
This work builds upon the assumption that the nucleon wave
function contains an admixture of negatively polarized $\bar ss$
quark pairs and allows one to deduce some qualitative predictions
for the state dependence of $\phi$ production in $\bar NN$,
but also in $NN$ collisions too. Specifically, this picture has been used
to explain the observed differences in the branching ratios for
$\bar pp \rightarrow \phi \pi^0$ at rest - which is rather large
and OZI violating when the initial $\bar pp$ system is in a $^3S_1$
state, but small and consistent with the OZI rule for the $^1P_1$
initial state. With regard to $\phi$ production in $NN$ collisions,
it is argued that $\phi$ production should be likewise much larger 
from initial spin-triplet states \cite{Ell99}. 
Specifically, this means that the $\phi$ production cross section 
in the $pp$ induced reaction is expected to be enhanced
over the OZI rule because, close to threshold, it will 
proceed predominantly via the transition 
$^3P_1\rightarrow ^1S_0 s$. (We use here the standard nomenclature 
for labeling the partial waves of the $NN$ states and the angular
orbital momentum of the $\phi$ relative to the final $NN$ system.) 
On the other hand, for $pn$ induced reactions like $pn \rightarrow d\phi$,
the production amplitude is given by the transition
$^1P_1\rightarrow ^3S_1 s$ and should therefore be OZI suppressed.
In contrast, our dynamical model predicts cross section ratios of
comparable magnitude for the $pp \rightarrow ppv$ and 
$pn \rightarrow dv$ reactions, as can be seen from a comparison of 
Figs.~\ref{stratio} and \ref{ppratio}. Clearly, experimental 
information on these ratios would be very useful.

\section{Summary}

We have investigated the reactions $pn\rightarrow d\phi$ and
$pn\rightarrow d\omega$ near threshold in a relativistic 
meson-exchange model. The elementary vector-meson production
mechanism is provided by the nucleonic and the $\phi\rho\pi$
(and $\omega\rho\pi$, respectively) meson-exchange currents. 
The presented results for total cross
sections and for the angular distributions of the vector
mesons are true predictions. All model parameters have been
taken from a previous study of the reactions 
$pp\rightarrow pp\phi$ and $pp\rightarrow pp\omega$ \cite{Nak2}.

A clear consequence of the model is 
that the reaction $pn\rightarrow d\phi$ is almost completely
dominated by the meson exchange current. The contributions from
the nucleonic current, which were found to be already small for
$pp\rightarrow pp\phi$, are now even less significant - especially
for the total cross section. The angular distribution of the 
$\phi$ meson, however, still reflects some influence of the nucleonic
current through interference effects. It is predicted to be rather
flat for excess energies $Q \lsim$ 30 MeV but to show an increasing
suppression of very forward and backward angles for higher energies. 

In case of the reaction $pn\rightarrow d\omega$ the contributions
from the nucleonic current are somewhat larger. Thus, stronger 
interference effects between this current and the meson-exchange 
current are possible. As a consequence, the predictions 
for $pn\rightarrow d\omega$ based on the four parameter sets of 
Ref.~\cite{Nak2} differ much more than those for
$pn\rightarrow d\phi$, showing variations by up to a 
factor 3 in the total cross sections. This also implies greater
variations in the cross section ratio 
$\sigma_{pn\rightarrow d\phi} / \sigma_{pn\rightarrow d\omega}$, 
for which our model yields values that exceed the estimate
based on the naive OZI rule by factors of up to 8. 

\acknowledgements{We would like to thank J. Durso for a careful
reading of the manuscript.} 

\vfill \eject

\newpage

\begin{table}
\caption{Parameters of the four different models employed in
the present study. The values 
$g_{\phi\rho\pi}$ = -1.64, $g_{\omega\rho\pi}$ = 10, 
$g_{NN\omega}$ = 9, and $\Lambda_M$ = 1450 MeV are used for
all four sets. $\kappa \equiv \kappa_v = f_{NNv}/g_{NNv}$ with 
$v = \phi,\omega$.}
\vskip -0.25cm
\tabskip=1em plus2em minus.5em
\halign to \hsize{\hfil#\hfil&\hfil#\hfil&\hfil#\hfil&
                  \hfil#\hfil&\hfil#\hfil\cr
\noalign{\hrulefill}
Set  &   1  &  2  &   3  &   4  \cr
\noalign{\hrulefill} 
$\kappa  $ & -0.5  & -0.5 & +0.5 & +0.5             \cr
$\Lambda_N$   &  1170   &   1411  &  1312   &  1545    \cr
$g_{NN\phi}$  &  -0.45  &  -0.19  &  -0.90  &  -0.40   \cr
\noalign{\hrulefill} }
\label{tab1}
\end{table}

\begin{table}
\begin{center}
\caption{Total cross sections for the reactions $pn\rightarrow d\omega$
and $pn\rightarrow d\phi$ resulting from the model parameter sets 
$1$ - $4$ at selected excess energies $Q$.}
\label{tab2}
\begin{tabular}{|c||c|c|c|c||c|c|c|c|}
&\multicolumn{4}{c||}{$pn\rightarrow d\phi$}&
\multicolumn{4}{c|}{$pn\rightarrow d\omega$}\\
&\multicolumn{4}{c||}{$\sigma_{tot}$ (nb) }&
\multicolumn{4}{c|}{$\sigma_{tot}$ ($\mu$b) }\\
\hline
Set  &   1  &  2  &   3  &   4   &   1  &  2  &   3  &   4  \\
\hline
$Q$ = 30 MeV & 219.9 & 216.4 & 265.8 & 253.1  & 19.2 & 9.4 & 31.1 & 26.9   \\
$Q$ = 60 MeV & 283.8 & 279.6 & 341.3 & 323.8  & 27.2 & 17.7 & 41.8 & 38.9   \\
$Q$ = 100 MeV & 322.4 & 318.2 & 385.4 & 364.2  & 34.1 & 26.9 & 49.9 & 50.2   \\
\end{tabular}
\end{center}
\end{table}

\newpage

\begin{figure}
\caption{$\phi$ and $\omega$-meson production currents, $J^\mu$, included in 
the present study: (a) nucleonic current, (b) $v\rho\pi$ meson exchange 
current. $M = \pi, \eta, \rho, \omega, \sigma, a_o$.}
\label{fig1}
\end{figure}

\begin{figure}
\caption{Total cross section for the reaction $pn\rightarrow d\phi$ as
predicted by the model set 1 of Table~\ref{tab1} as a function of excess 
energy $Q$. The solid line is
the result of the full model. The dashed (dash-dotted) curves show the 
contributions of the nucleonic current (meson exchange current) alone.}
\label{stphi}
\end{figure}

\begin{figure}
\caption{Angular distribution in the center-of-mass system of the produced 
$\phi$-meson in the reaction $pn\rightarrow d\phi$ at the excess energies 
$Q$ = 30 MeV (left panel) and 100 MeV (right panel).
The dashed, solid, dashed-dotted and dotted curves are the 
results for the model parameter sets 1, 2, 3, and 4.}
\label{sdphi}
\end{figure}

\begin{figure}
\caption{Same as Fig.~\ref{stphi} for the reaction $pn\rightarrow d\omega$ as
predicted by the model sets 1 (a), 2 (b), 3 (c), and 4 (d) of Table~\ref{tab1}.
The dotted curves correspond to the results with the sharp mass of $728.6 MeV$ 
for the $\omega$-meson.}
\label{stome}
\end{figure}

\begin{figure}
\caption{Same as Fig.~\ref{sdphi} for the reaction $pn\rightarrow d\omega$.} 
\label{sdome}
\end{figure}

\begin{figure}
\caption{Ratio of the total cross sections $\sigma_{pn\rightarrow d\phi} /
\sigma_{pn\rightarrow d\omega}$ as a function of excess energy $Q$. 
The dashed, solid, dashed-dotted and dotted curves are the results 
for the model parameter sets 1, 2, 3, and 4. The horizontal solid line 
indicates the value predicted by the naive OZI rule.}
\label{stratio}
\end{figure}

\begin{figure}
\caption{Same as Fig.~\ref{stratio} for the ratio
$\sigma_{pp\rightarrow pp\phi} / \sigma_{pp\rightarrow pp\omega}$.}
\label{ppratio}
\end{figure}

\newpage 

\vglue 1cm 
\begin{center}
\begin{figure}
\epsfig{file=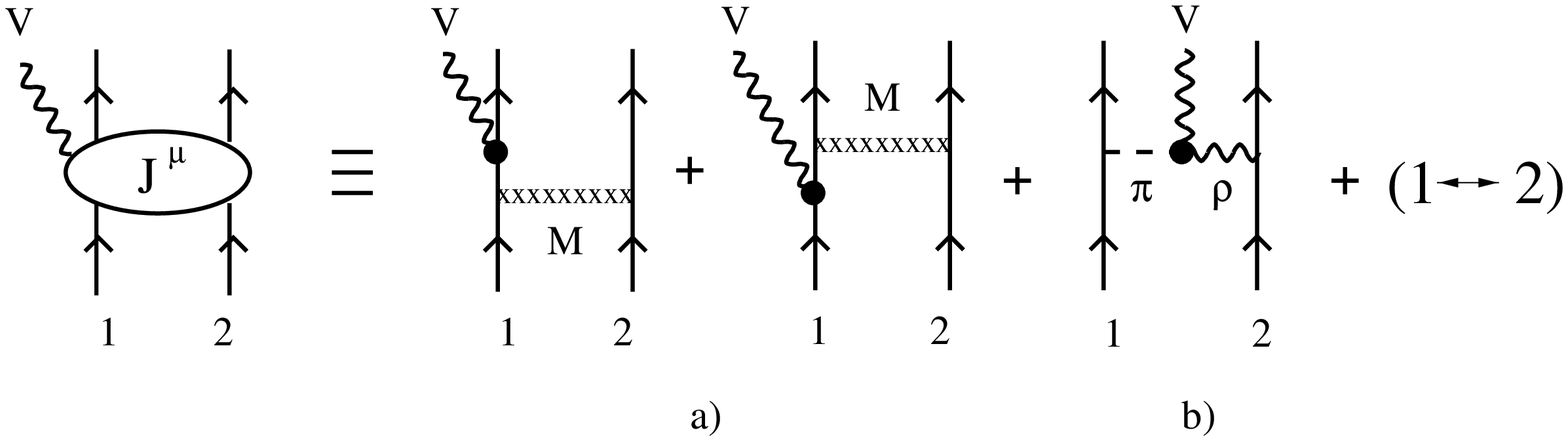, width=15.0cm} 
\end{figure}
\end{center}

\vskip 4cm 

\center{Fig. 1} 

\newpage 
\vglue 1cm 

\begin{center}
\begin{figure}
\epsfig{file=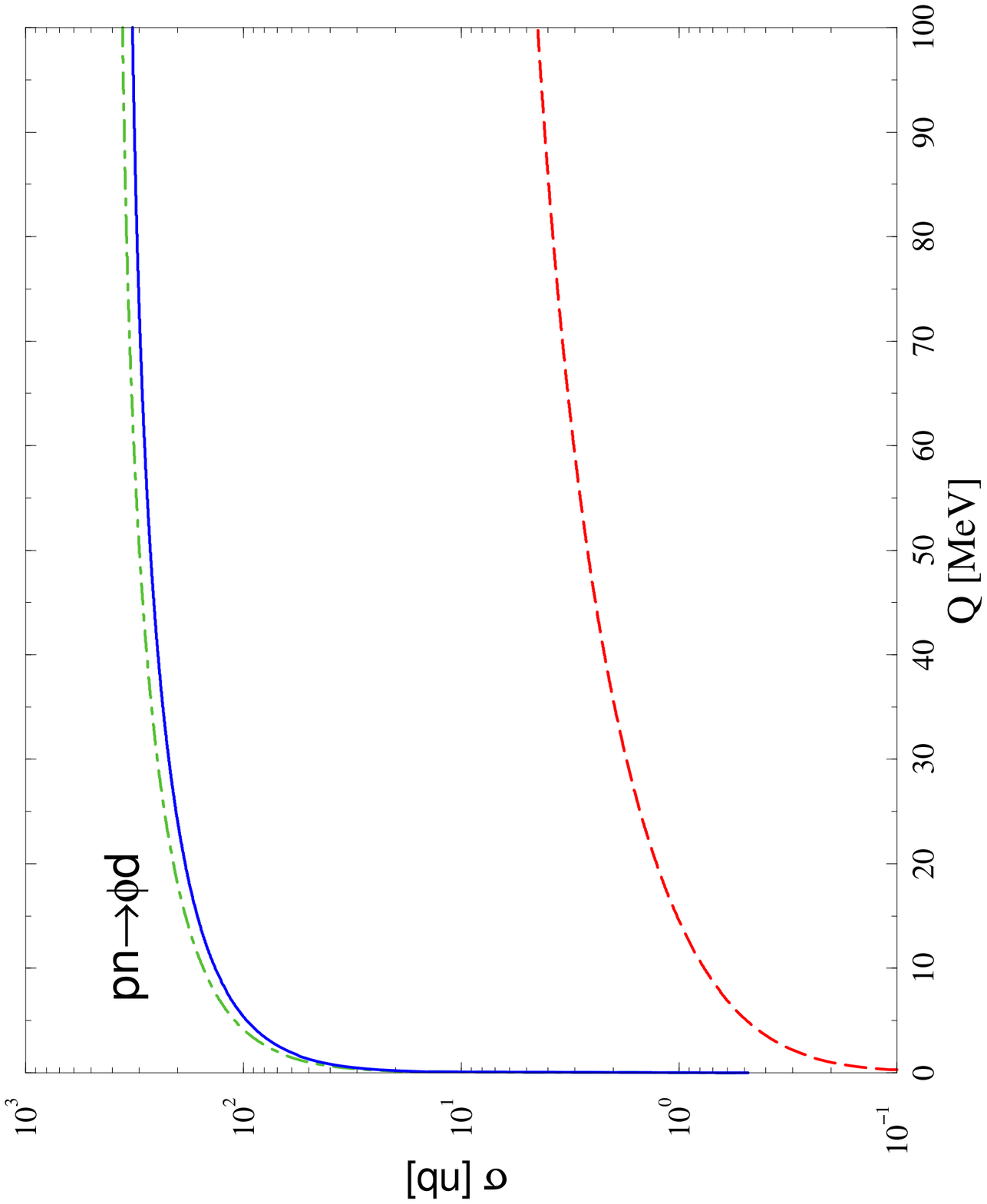, width=15.0cm, height=15.0cm} 
\end{figure}
\end{center}

\vskip 2cm 

\center{Fig. 2} 

\newpage 
\vglue 1cm 

\begin{center}
\begin{figure}
\epsfig{file=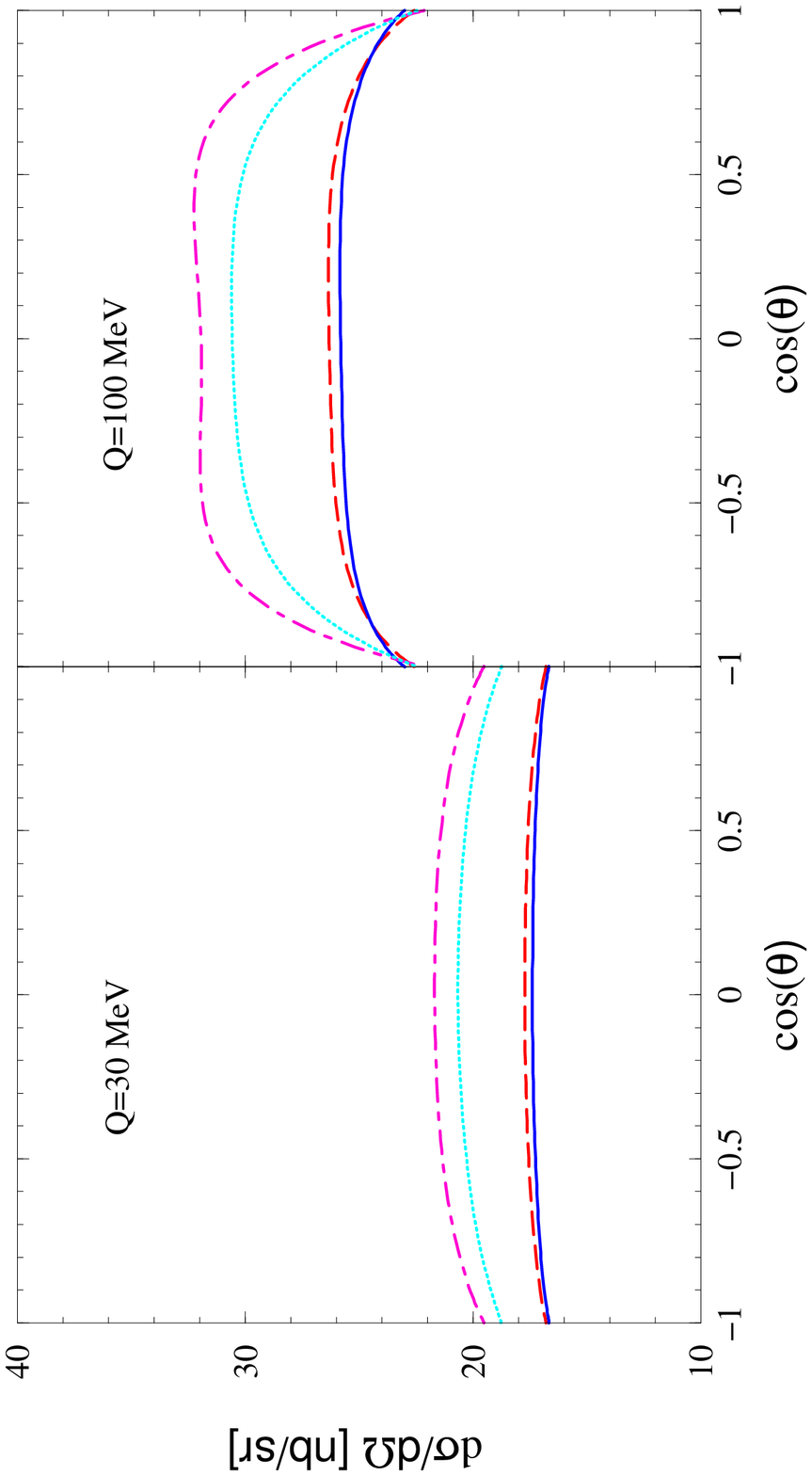, width=10.0cm, height=15.0cm} 
\end{figure}
\end{center}

\vskip 2cm 

\center{Fig. 3} 

\newpage 
\vglue 1cm 

\begin{center}
\begin{figure}
\epsfig{file=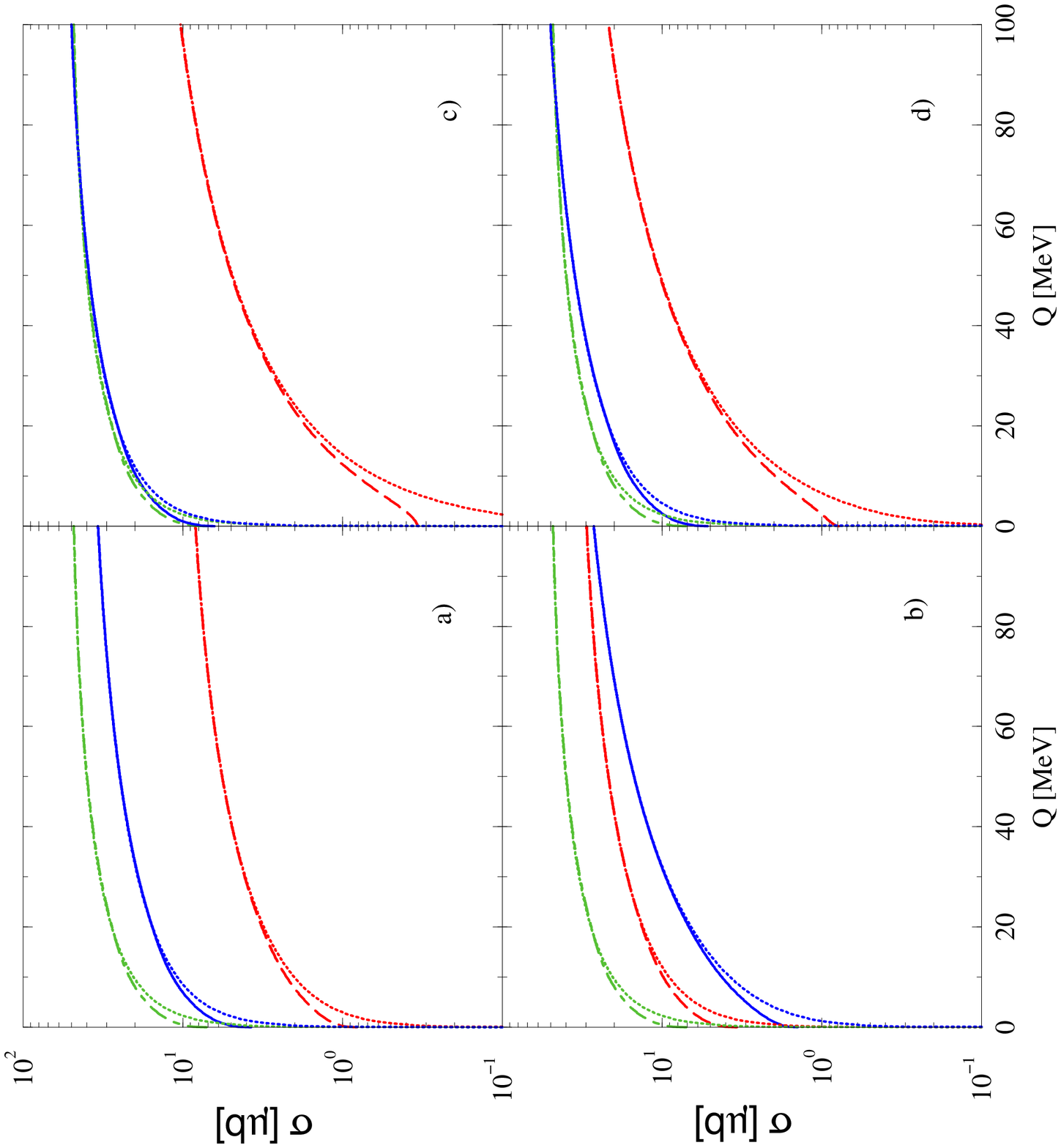, width=15.0cm, height=15.0cm} 
\end{figure}
\end{center}

\vskip 2cm 

\center{Fig. 4} 

\newpage 
\vglue 1cm 

\begin{center}
\begin{figure}
\epsfig{file=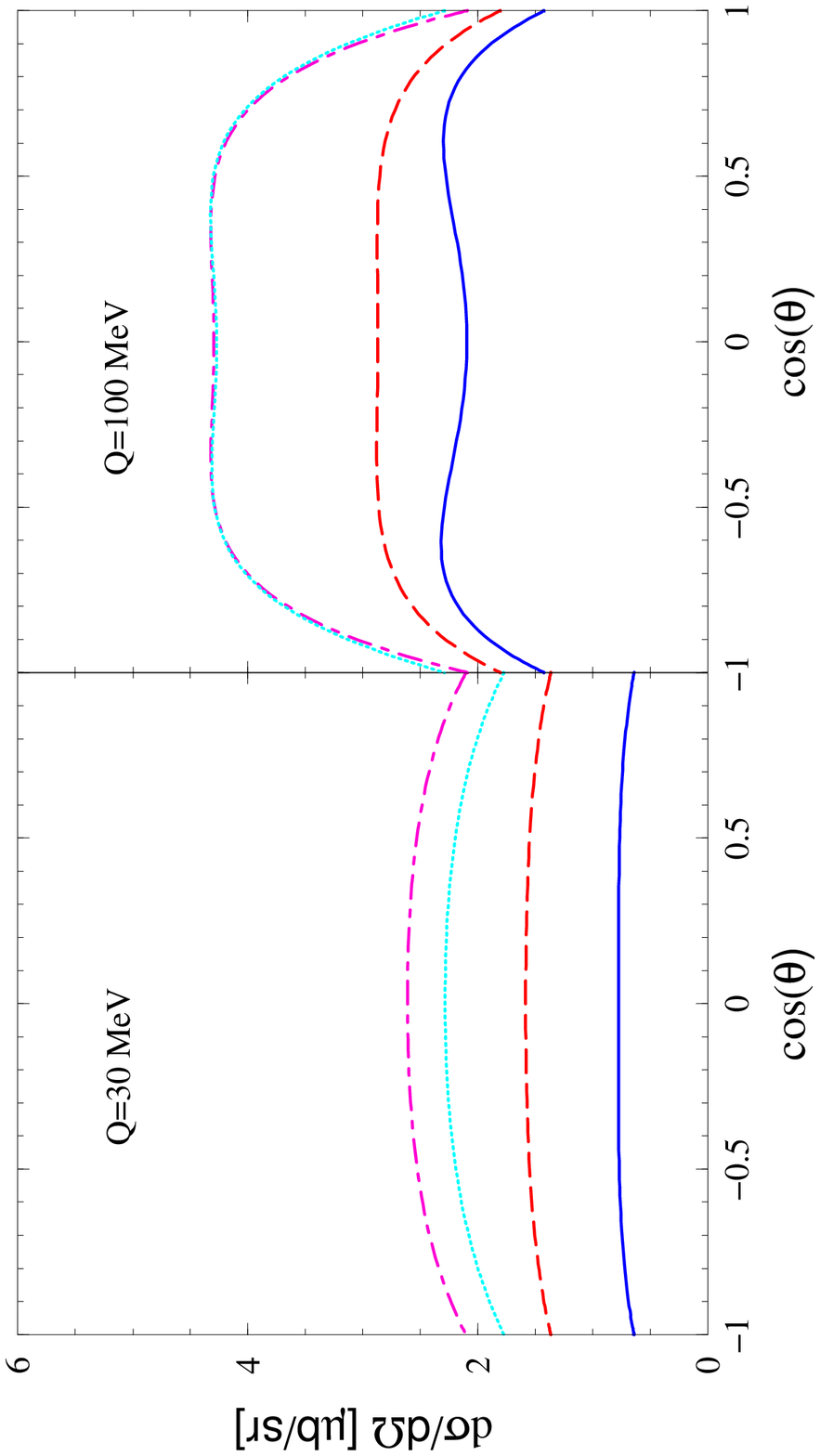, width=10.0cm, height=15.0cm} 
\end{figure}
\end{center}

\vskip 2cm 

\center{Fig. 5} 

\newpage 
\vglue 1cm 

\begin{center}
\begin{figure}
\epsfig{file=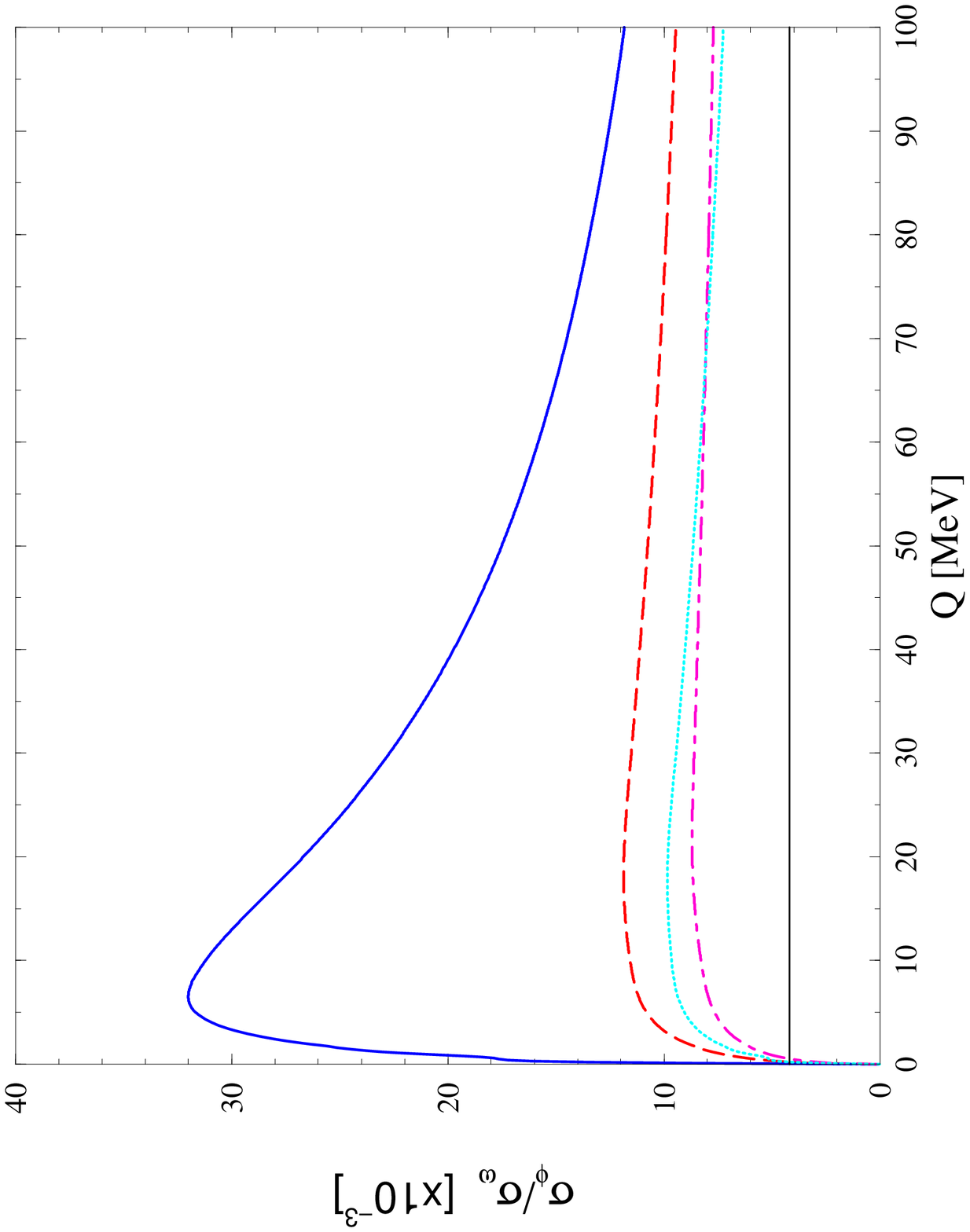, width=15.0cm, height=15.0cm} 
\end{figure}
\end{center}

\vskip 2cm 

\center{Fig. 6}

\newpage 
\vglue 1cm 

\begin{center}
\begin{figure}
\epsfig{file=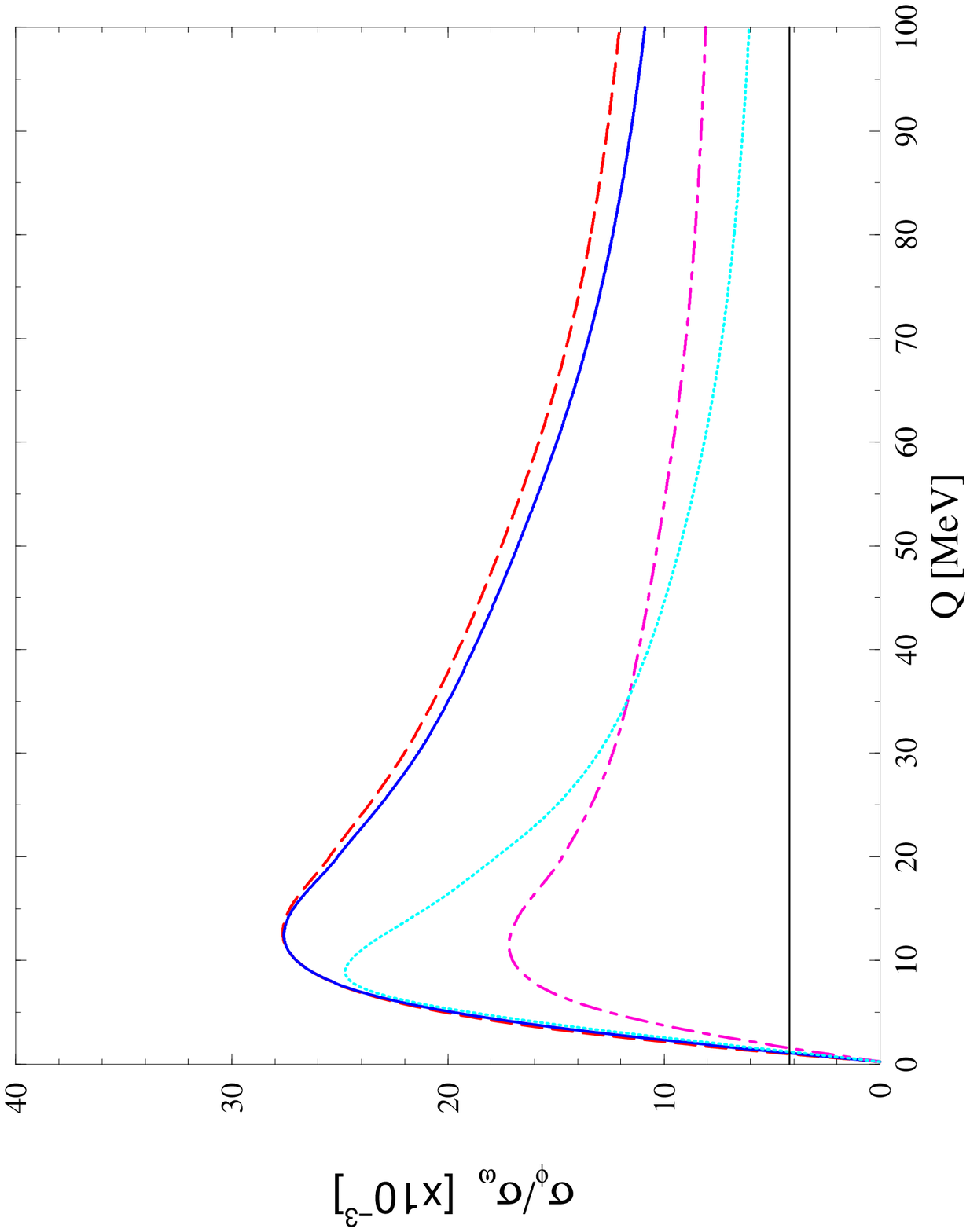, width=15.0cm, height=15.0cm} 
\end{figure}
\end{center}

\vskip 2cm 

\center{Fig. 7} 

\end{document}